\documentclass[twocolumn,letter]{jpsj2} 
\def\simge{\lower0.7ex\hbox{$\ \overset{>}{\sim}\ $}}
\def\simle{\lower0.7ex\hbox{$\ \overset{<}{\sim}\ $}}

\title{Phonon Dynamics and Multipolar Isomorphic
Transition in
$\beta$-pyrochlore KOs$_2$O$_6$}

\author{Kazumasa HATTORI and Hirokazu TSUNETSUGU}

\inst{Institute for Solid State Physics, University of Tokyo, 5-1-5, Kashiwanoha, Kashiwa Chiba 277-8581, Japan
}

\abst{We investigate with a microscopic model anharmonic K-cation 
oscillation observed by neutron experiments in $\beta$-pyrochlore 
superconductor KOs$_2$O$_6$, which also shows a mysterious first-order 
structural transition at $T_p$=7.5 K. 
We have identified a set of microscopic model parameters that 
successfully reproduce the observed temperature dependence and the 
superconducting transition temperature. 
Considering changes in the parameters at $T_p$, we can explain 
puzzling experimental results about electron-phonon coupling and 
neutron data. Our analysis demonstrates that the first-order transition
is multipolar transition driven by the octupolar component of K-cation 
oscillations. The octupole moment does not change the symmetry and is 
characteristic to noncentrosymmetric K-cation potential. }

\kword{beta pyrochlore, anharmonic phonon, superconductivity, isomorphic
transition}

\begin{document}
\maketitle

The $\beta$-pyrochlore oxides $A$Os$_2$O$_6$($A$=K, Rb, or Cs) 
are one of the material families that have unique cage-like structure,
and $A$-cation inclusions exhibit large anharmonic local oscillations\cite{HiroiUnprecedented,Hiroi2ndAno,ThermalCondKasahara,SummaryHiroi,Yoshida,IsotropicGappSCSC,peneShimono,Hasegawa,
Nagao}. Such anharmonic and large-amplitude oscillations 
strongly interact with conduction electrons, which leads to strong
 coupling superconductivity\cite{BattloggSC,Manalo,Chang}, anomalous
 temperature dependence of electric resistivity\cite{SummaryHiroi,MahanSofo} and
 nuclear relaxation time\cite{DahmUeda,Yoshida}. To explain such
 interesting properties is a big challenge for the theory of electron-phonon 
 systems and strong coupling theory of superconductivity\cite{Eliashberg}.

As the $A$-cation size decreases, amplitude of the $A$-cation oscillation increases,  and
thus potassium oscillations are the most anharmonic\cite{Yamaura}. 
Elastic neutron scattering data show that the temperature
dependence curve of oscillation amplitude is concave (convex upwards) in KOs$_2$O$_6$, 
reflecting strong anharmonicity\cite{NeutronSasai,Galati}.
A recent inelastic neutron scattering experiment
revealed  that the phonon
peaks shift to lower energy with decreasing temperature and 
 the peak positions  are at around 3-7
meV at the lowest temperature\cite{Mutka}. 
This softening is the strongest in K-, and the weakest in Cs-compound.
Superconducting transition takes places in all the three, and the
transition temperature is $T_c=$9.6, 6.3 and 3.3 K for $A$=K, Rb and Cs,
respectively\cite{SummaryHiroi,Nagao}. The symmetry of the 
superconducting gap function is confirmed to be {\it s}-wave\cite{IsotropicGappSCSC,peneShimono}.

KOs$_2$O$_6$, the most anharmonic one, not only has the highest $T_c$ of
superconductivity, but also exhibits another singularity, a 
first-order structural transition at
$T_p=7.5$ K\cite{Hiroi2ndAno}. It is interesting that no 
sign of symmetry breaking has been
observed\cite{Hasegawa,newSasai,Yamaura2}. Below $T_p$, the temperature dependence of electric
resistivity changes to 
$T^2$ from $T^{\gamma}(\gamma\sim 0.5)$
 at higher temperatures\cite{SummaryHiroi}, 
and the absolute
value decreases by about 25 \% at $T_p$.
Both of these results show
that electron-phonon scatterings are suppressed
 below $T_p$\cite{ThermalCondKasahara}, which is also supported by the
 reduction in the specific heat jump at $T_c$ in magnetic fields $H$
 when $T_c(H)<T_p(H)\sim T_p(0)$\cite{SummaryHiroi}. However, recent
 neutron scattering data\cite{newSasai} show the oscillation amplitude 
 increases below $T_p$. Its naive interpretation is an increase in 
the electron-phonon coupling.


\begin{figure}[b]
\vspace{-0.5cm}
\begin{center}
    \includegraphics[width=0.46\textwidth]{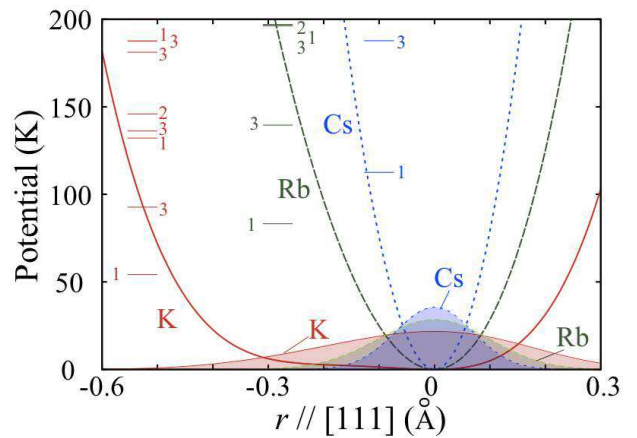}
\end{center}
\caption{(color online) Potential profile,  and probability
 density of the ground state wavefunction $|\Psi_{\rm gs}({\bf r})|^2$
 for K (solid), Rb (dashed), and Cs (dotted line),
 respectively, along [111] direction. Energy eigenvalues are indicated by
 short bars and the numbers represent the degeneracy.
 }
\label{fig-Pot}
\vspace{-0.5cm}
\end{figure}

In our previous work\cite{HatPRB}, we have developed a general theory of 
anharmonic ion oscillations in tetrahedral symmetry. 
A strong coupling theory of superconductivity mediated by these ion
oscillations has been also developed. Applying them to
$\beta$-pyrochlore compounds, 
we estimated their $T_c$. We also started to study the contradiction
 between the change of the electron phonon coupling and the
neutron data at $T_p$ in KOs$_2$O$_6$ but that was in a qualitative level. 

In this Letter, we will resolve by microscopic calculations 
 the contradiction between the data of resistivity and specific heat \cite{SummaryHiroi} and the
 neutron scattering\cite{newSasai} in KOs$_2$O$_6$ at $T_p$. 
Our calculation demonstrates that the first-order transition is 
{\it a multipolar isomorphic phase transition} and the phonon
anharmonicity plays a crucial role to explain the experimental data.
We will apply the theory developed in ref. 
 22 to $A$Os$_2$O$_6$ and carry out more elaborate
numerical calculations in order to understand the higher-temperature and
 higher-energy properties of the anharmonic $A$-cation oscillations.
We will show that we can reproduce the temperature dependence of the softening of phonon energy and the amplitude of $A$-cation
oscillations in a wide range of temperatures in nice agreement with the experimental data.

We start with a short review of our model for $A$-cation oscillation. 
Since the inelastic neutron experiment shows 
nearly momentum-independent modes of their dynamics\cite{Mutka}, 
we employ a local model. 
The $A$-site has the tetrahedral $T_d$ 
symmetry and this implies that the ion potential
generally has, in addition to spherical and cubic fourth order terms, 
 a third-order anharmonic term, which
breaks space inversion symmetry. The ionic Hamiltonian is thus given by
\begin{eqnarray}
H_{\rm ion}\!&=&\!-\frac{\nabla^2_{\bf r}}{2M}+\frac{M\omega^2}{2}|{\bf r}|^2
         +bxyz
+c_{1}|{\bf r}|^4+c_{2} \tilde{r}^4,\label{Hph}
\end{eqnarray}
where $\hbar$ is set to be unity, and 
${\bf r}=(x,y,z)$ is the ion displacement from
 the equilibrium position, $\tilde{r}^4=x^4+y^4+z^4$ and $M$ is the ion
 mass. 
We ignore the higher order terms
 of $O(|{\bf r}|^5)$ which is irrelevant for our discussions because of
 positive forth order terms. 
We diagonalize Hamiltonian (\ref{Hph}) numerically
 in the restricted Hilbert space of dimension 62196, which is about
 three times larger Hilbert space used in the previous
 study\cite{HatPRB} and sufficient to
 discuss the physical properties below room temperature, 
by extrapolating the obtained data to the infinite limit if necessary.
Details of calculations are explained in
 ref. 22. 

An important point is the variation in potential parameters among
different compounds. 
The Madelung part of the potential energy 
is essentially the same
among the three compounds, since the band structure calculations\cite{BandCal} show
essentially the same electric states for them.
Thus, the main difference in the
potential parameters originates from the
the relaxation of local charge density as discussed in ref. 23. 
This point can be well captured by setting the 
smaller $\omega$ for the smaller ion.  
We use the same parameters as those in ref. 22:
$\omega=26.4$ K, $54.6$ K and
$74.8$ K 
 for K, Rb, and Cs, respectively, with keeping the same
values for other parameters, $b=9324$ K/\AA$^3$ and $c_{1}=4c_{2}=3332$ K/\AA$^4$. 
The validity of our choice will be checked later by comparing 
the phonon energy and oscillation amplitudes with experimental data.


Figure \ref{fig-Pot} shows the potential profile with these parameters
 along [111] direction
for the three compounds. Eigenenergies of 
 Hamiltonian (\ref{Hph}) are
 indicated by bars and the probability density of the ground state 
$|\Psi_{\rm gs}({\bf r})|^2$ is also shown. 
The potential of K-cation is 
much shallower than those of  Rb and Cs, and correspondingly, 
its $|\Psi_{\rm gs}({\bf r})|^2$ has a broader tail than 
 the others. Kune\v{s} {\it et al.,}
proposed a potential with $\omega^2<0$ for 
KOs$_2$O$_6$ based on the first principle
 calculation\cite{BandCal}, 
but with their choice, the K-cation oscillates about 1 \AA,  
which is too large and inconsistent with the neutron experiment
 data\cite{NeutronSasai,Galati}. Therefore, we use positive $\omega^2$ for
 all the three compounds.
As we will show later, 
our potential parameters reproduce the correct temperature dependence of the  
variance of ion oscillation $\langle x^2\rangle$, 
and also the excitation
energy agreeing with the neutron and x-ray scattering 
data\cite{Yamaura, Mutka, NeutronSasai, Galati}. 

\begin{figure}[tb]
\begin{center}
    \includegraphics[width=0.46\textwidth]{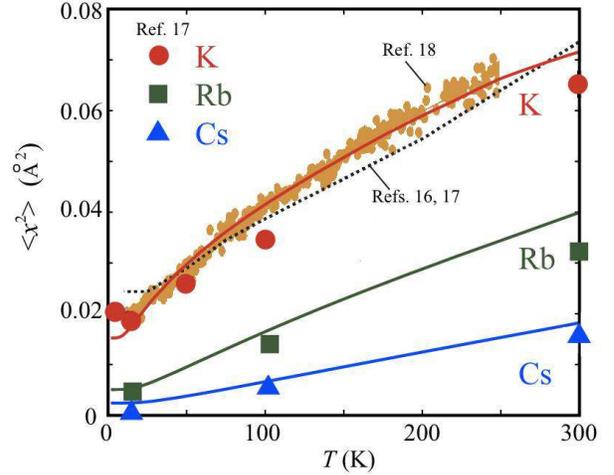}
\end{center}
\vspace{-0.5cm}
\caption{(color online) Temperature dependence of the variance of ion
 oscillation $\langle x^2
 \rangle$. Three solid lines represent calculated $\langle
 x^2\rangle$ for the three compounds. Symbols and dotted line show the 
experimental data taken from
 refs. 16, 17 and 18. }
\label{fig-DW}
\vspace{-0.5cm}
\end{figure}

We first calculated the variance of ion oscillation $\langle x^2\rangle$
as a function of temperature from the calculated eigenfunctions of
(\ref{Hph}) and the results are shown in Fig. \ref{fig-DW}.
The experimental values determined by neutron\cite{NeutronSasai,Galati} and
x-ray\cite{Yamaura,NeutronSasai} 
scattering are also shown for comparison. 
 As is clearly seen,
our results are quantitatively consistent with the experimental
results. For KOs$_2$O$_6$, we mainly concentrate on fitting the recent 
experimental data of ref. 18 and the low-temperature limit is
consistent with the recent data $\sim 0.014$ \AA$^2$ just above $T_p$\cite{newSasai}. 
As for Rb- and
Cs-compounds, the calculated $\langle x^2\rangle$ is
slightly larger than the data in ref. 17. However, concerning
the discrepancy between the data in refs. 17 and 18 for
K-compound, our results for Rb- and Cs-compounds well capture the
essential aspect of the experimental data. 
We note that the 
values in the  low-$T$ limit are important to discuss the
superconducting transition temperature, 
since these values directly 
influence the dimensionless electron-phonon coupling constant $\lambda$.

\begin{figure}[tb]
\begin{center}
    \includegraphics[width=0.46\textwidth]{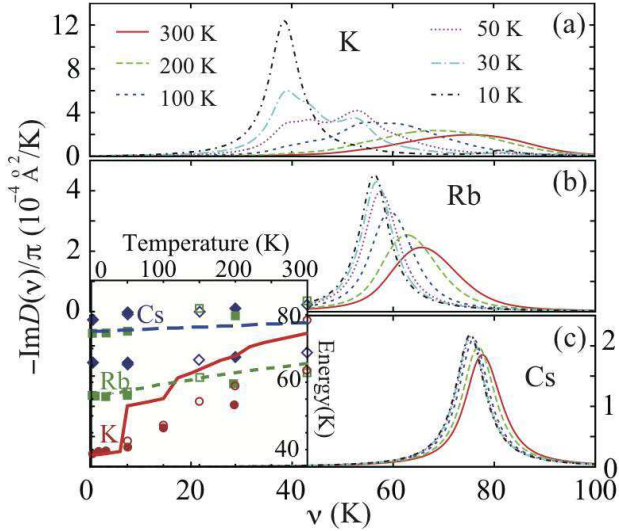}
\end{center}
\vspace{-0.5cm}
\caption{(color online) Phonon spectrum vs energy at six different
 temperatures for (a) K, (b) Rb, and (c) Cs compounds. 
Spectrum is broaden owing to the many Lorentzians whose width is given 
by $\eta=3.5$
 K and this is the direct consequence of the anharmonicity.
Inset: Temperature dependence of the highest peak position for K
 (circle), Rb(square) and Cs(diamond) taken from ref. 19. Lines represent
 the present calculations. 
} 
\label{fig-DOS}
\vspace{-0.6cm}
\end{figure}

Figure \ref{fig-DOS} shows the phonon spectrum, i.e., the
imaginary part of the phonon Green's function $D(\nu)=-i\int_0^{\infty}
dt\ 
e^{i\nu t-\eta t}\langle [x(t),x(0)]\rangle$ at six different 
temperatures for $A$=K, Rb and Cs compounds. $\eta$ is a
phenomenological relaxation rate
 set as $\eta=3.5$ K. There are several peaks in the spectrum of 
KOs$_2$O$_6$ for intermediate temperatures in Fig. \ref{fig-DOS}(a) 
owing to the anharmonic terms
 and the smaller $\omega$ compared to other members. 
The results are consistent with the  
 inelastic 
neutron scattering spectra\cite{Mutka}, particularly in the
following three aspects. 
(i) The peak
positions at the lowest temperature are 38.5, 52.9 and 73.1 K, which correspond to the
energy of the first excited states $\Delta$,
consistent with the neutron data at 1.5 K\cite{Mutka}: $\Omega_E^{\rm exp}=38$, 56, and 66 K for
K, Rb and Cs, respectively. 
(ii) The softening of the peak energy with decreasing $T$  is 
the strongest for K, intermediate for Rb and the weakest for Cs. 
The temperature dependence is also qualitatively reproduced by our
calculation as shown in the inset of Fig. \ref{fig-DOS}. Note that 
the relevant mode splits into two branches owing to two cages in the
unit cell, and our data correspond to their average energy. 
(iii) The spectrum of K at higher temperatures is much broader than that 
of Rb and Cs.

As discussed above, our model can 
explain the neutron experiments and well describes potential  
 of $A$-cation oscillations. These results confirm our previous work,\cite{HatPRB} in
 which the superconducting transition temperature $T_c$ was calculated
 with the same parameters used in this
 Letter and the result is $T_c=$ 10.5, 5.7 and 3.4 K, for K,
 Rb and Cs, respectively. The calculated $T_c$ agrees
 with the experimental data within $\pm 1$ K, even without
optimizing parameters for each compound. 
The variation in $T_c$ among the three members is 
mainly attributed to the difference in the anharmonicity of $A$-cation 
oscillations. Indeed, 60 \% of $T_c$ is owing to the K-cation
oscillation in KOs$_2$O$_6$.


As mentioned in the introduction, KOs$_2$O$_6$ 
exhibits a first-order structural transition at $T_p=7.5$ K, which 
 does not break any symmetry\cite{SummaryHiroi,Hasegawa,newSasai,Yamaura2}. 
In ref. 24, we
have proposed a scenario of isomorphic structural 
transition to explain it based on a simple toy model.  
The point is that in the tetrahedral symmetry, not only 
$\langle {\bf r}^2 \rangle=3\langle x^2\rangle\equiv 3u_2$ but 
also the third moment $|\langle xyz \rangle|\equiv u_3$ are nonzero and
their changes at $T_p$ do not break the point group symmetry.

Here, we will examine this scenario by means of the present realistic 
model. Analyzing the changes in the oscillation profile as a 
function of variations in the potential parameters at $T_p$, we will
identify the transition as  
 ``{\it multipolar}'' isomorphic one,  
because the change in the octupole $u_3$ 
is much larger than in the isotropic scalar $u_2$. 
This point is important to explain the
experimental results of recent neutron scattering\cite{newSasai} and
specific heat jump at $T_c$\cite{SummaryHiroi}, 
which will be explained below.  
Our previous analysis based on the simplified toy
model\cite{Hat} fails to describe this point, since the model cannot  
distinguish $u_2$ and $u_3$ owing to the lack of the degrees of freedom. 
This kind of phase transition driven by 
a scalar order parameter is similar to that proposed for the f-electron
system PrRu$_4$P$_{12}$\cite{Takimoto}. 


\begin{figure}[!t]
\begin{center}
    \includegraphics[width=0.5\textwidth]{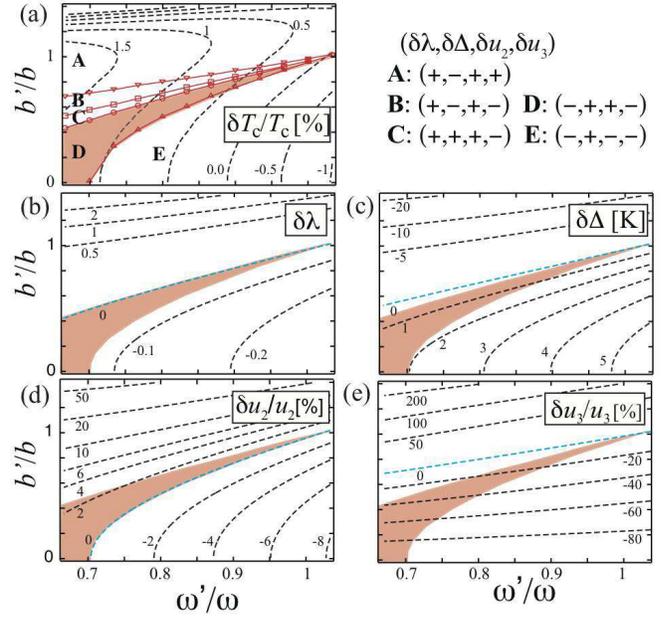}
\end{center}
\vspace{-0.5cm}
\caption{(color online) Contour plot of (a) $\delta T_c$ in
 $\omega'$-$b'$ plane with fixing $c'_1$$=$$5c'_2$$=$$c_{1}$, 
 (b) $\delta\lambda$, (c) $\delta\Delta$, (d) $\delta u_2/u_{2}$, and
 (e) $\delta u_3/u_{3}$. In (a), all the five regions are indicated; 
A: $({\rm sgn}(\delta\lambda),\ {\rm sgn}(\delta\Delta),\ {\rm
 sgn}(\delta u_2),\ {\rm sgn}(\delta u_3))=(+,-,+,+)$, B: $(+,-,+,-)$,
 C: $(+,+,+,-)$,
 D: $(-,+,+,-)$ and E: $(-,+,-,-)$. The solid lines with symbols represent
 the boundaries and the dashed lines represent the contours. 
Region D is indicated by shade.}
\label{fig-isomor}
\vspace{-0.5cm}
\end{figure}

Let us discuss the nature of the 
transition at $T_p$ and three experimental results are important. First, the magnetic field dependence of the
specific heat jump at $T_c$\cite{SummaryHiroi} 
indicates that the electron-phonon coupling $\lambda$ decreases below $T_p$.
Secondly, a recent neutron 
experiment shows that $u_2$ jumps up at $T_p$ and 
the value is $\sim 0.02$ \AA$^2$ at low temperature 1.5
K\cite{newSasai}. 
Thirdly, the upper critical field $H_{c2}(T)$ shows that 
the first-order transition does not
affect $T_c(H\to 0)$\cite{SummaryHiroi, explainHc2}.
The first two experimental data cannot be explained on the basis of harmonic
oscillations. In the harmonic case, $\lambda\propto u_2^2$ and thus,
the increase in $u_2$ leads to enhanced $\lambda$. 
These puzzling 
experimental results are naturally
explained on the basis of the multipolar isomorphic transition.

Since the transition is first-order, 
the potential parameters change discontinuously, which originates from, 
 for example, the changes in volume, oxygen positions, and
the inter-site ion interactions. 
The high-$T$ potential parameters 
 $\omega$$=$$26.4$ K, $b$$=$$9324$ K/\AA$^3$, 
$c_{1}$$=$$4c_{2}$$=3332$ K/\AA$^4$ change to low-$T$ ones, say, $\omega'$, $b'$, $c'_1$
and $c'_2$, respectively. 
Changes in $b$ and $\omega$ are particularly important 
and we investigate the effects of their change, while we set 
$c'_1=$$5c'_2$$=c_{1}$ for simplicity. Thus, with varying $\omega'$ and $b'$, 
we repeat the same procedure as before\cite{HatPRB} and calculate the
deviations of dimensionless electron-phonon 
coupling $\delta \lambda$$\equiv$$\lambda'$$-\lambda$, and gap 
$\delta\Delta$, similarly, $\delta u_2$, $\delta u_3$, and $\delta T_c$. 
  Here, the quantities denoted with prime symbol 
 are the low-$T$ values calculated for the new potential parameters at
 $T_c$ for each of the parameter set. $T_c$ is low enough compared with 
$\Delta$ and thus these quantities are regarded as those 
for the low-temperature limit.
 The results are shown in Fig. \ref{fig-isomor}.

%
%
First, it is noted that there exist five regions, A-E, distinguished by the sign
 of $\delta\lambda$, $\delta\Delta$, $\delta u_2$ and $\delta u_3$. 
In the harmonic case, only the variations characterized by the region 
A or E are possible, but they are not consistent with the experimental
 results, $\delta\lambda <0$ and $\delta u_2 >0$.
The other three regions appear as a consequence of the third-order phonon
 anharmonicity $b$. 
Among them, it is the region D that satisfies the experimental constraints,  
and the low-$T$ parameters  $b'$ and $\omega'$ 
 should be located inside this region.
The change in $T_c$ is less than 2 \% and this small change is
 consistent with the last one of the three experimental points mentioned
 before. 

%
%

The puzzling behavior, $\delta \lambda<0$ and $\delta u_2 >0$, is a
consequence of two competing effects. The reduction in $\omega$ 
enhances $\lambda$, $u_2$, and $u_3$, whereas the reduction in $|b|$ 
suppresses them. The change in $\lambda$ is dominated by the
effect of reduced $|b|$, while the change in $u_2$ is rather due to 
 reduced $\omega$. Our calculation shows that the experimental 
results predict the constraint on the potential changes at $T_p$ as 
$2\delta \omega/\omega \simle \delta b/b \simle 1.5\delta
\omega/\omega<0$, as far as the changes are not so large. 
%
%

One can also explain the reduction 
in the electrical resistivity observed at $T_p$\cite{SummaryHiroi}, 
 since  the changes $\delta \Delta>0$ and  $\delta\lambda<0$ mean the
 suppression of the electron scatterings due to the K-cation oscillations. 
Finally, the nature of the first-order transition is identified by 
examining various moments of ion oscillations. 
As for the changes in $u_2$ and $u_3$,  the {\it isotropic} part $\langle {\bf r}^2\rangle=3u_2$ of the oscillation
 amplitude is enhanced 
while {\it anisotropic} one $|\langle xyz\rangle|=u_3$ is suppressed as shown in
Figs. \ref{fig-isomor}(d) and (e). Furthermore, the relative change in $u_3$ is much larger than 
that in $u_2$, and this is the reason to
call this a multipolar isomorphic transition.


The important point is that our calculation 
 already reproduces changes consistent with the experimental results at $T_p$.
 In order to carry out the further quantitative comparison between 
 the theory and the experimental results, 
one needs  more detailed information about the potential changes. 
It is also useful to observe the anisotropic part of the Debye-Waller
 factor and compare with theoretical calculation. 

In summary, 
we have proposed that the first-order
transition observed in KOs$_2$O$_6$ is a multipolar
isomorphic transition and demonstrated that the third order fluctuation 
$\langle xyz \rangle$ is the primary order parameter. Our theory
naturally explains both the reduction in electron-phonon coupling constant and the enhanced oscillation amplitude at $T_p$. 
We hope detailed neutron 
scattering experiments detect its change at $T_p$.



\section*{Acknowledgment}
The authors thank T. Dahm, K. Ueda, J. Yamaura and Z. Hiroi for
 discussions 
 and also acknowledge the international workshop ``{\it New
Developments in Theory of Superconductivity}'' held in Institute for
Solid State Physics, University of Tokyo, Japan, June 22-July 10, 2009, for
giving them an opportunity to discuss many aspects of this work. 
This work is supported by KAKENHI (No. 19052003 and No. 20740189).


\begin{thebibliography}{99}
\bibitem{explainHc2} $H_{c2}(T)$ below $T_p(H)$ for $H>8$ T can be
	extrapolated to vanishing at the same $T_c$ as the case of
	$H=0$, see Fig. 10 in ref. 4. 
        This implies that $T_c$ for the low-$T$ potential parameters is
	approximately the same as $T_c$ for the high-$T$ ones.
\end{thebibliography}
\end{document}